
\hcorrection{36pt}
\TagsOnRight
\magnification=\magstep1
\loadbold
\define\dr{\partial}
\define\dd#1{\partial_#1}
\define\du#1{\partial^#1}
\define\pos{\boldkey r}
\define\ndd#1#2{\eta_{#1#2}}
\define\nuu#1#2{\eta^{#1#2}}
\define\hdd#1#2{h_{#1#2}}
\define\huu#1#2{h^{#1#2}}
\define\hdu#1#2{h_#1{}^#2}
\define\drt#1#2{\frac{d#1}{d#2}}
\define\gdd#1#2{g_{#1#2}}
\define\guu#1#2{g^{#1#2}}
\define\ru{R{\cdot}u}
\define\rup#1{(R{\cdot}u)^#1}
\define\ra{R{\cdot}a}
\define\rap#1{R{\cdot}a)^#1}
\define\rw{R{\cdot}w}
\define\rwp#1{R{\cdot}w)^#1}
\define\rb{\overline r}
\define\aalign{\spreadlines{1\jot}\allowdisplaybreaks\align}
\define\tr{\text{tr\,}}
\documentstyle{amsppt}
\leftheadtext{Jaegu Kim}
\rightheadtext{Gravitational Field of a Moving Point Particle}
\topmatter
\title
Gravitational Field of a Moving Point Particle
\endtitle
\author
Jaegu Kim\footnote"*"{jaegukim\@cc\.kangwon\.ac\.kr}
\endauthor
\affil
Department of Physics,
Kangwon National University,
Chunchon 200--701 Korea
\endaffil
\abstract
The gravitational field of a moving point particle is obtained in
a Lorentz covariant form for both uncharged and charged cases. It is shown
that the general relativistic proper time interval at the location of the
particle is the same as the special relativistic one and the gravitational
and electromagnetic self forces are zero.
\flushpar
\endabstract
\endtopmatter
\document
\bigskip\flushpar
{\bf 1. Introduction}
\medskip\par
There has been a great deal of efforts to find general exact solutions to the
Einstein equations since the publication of general relativity. But as compared
to a simple linear equation in Newtonian theory, the Einstein equations are
very complicated coupled nonlinear equations. The metric tensor has 10
independent components instead of 1, and the Einstein equations are 10 coupled
equations of 8th degree rather than linear. Moreover, each equation has a
large number of terms which can hardly be handled by any analytic method.
Due to this complexity there is no general method of solving the equations.
However, the equations have been solved in a few cases in which the number of
variables is reduced by symmetry conditions on the metric tensor. The commonly
imposed symmetries are time independence, spherical symmetry and axial
symmetry, from which Schwarzschild solution [1], Reissner--Nordstr{\"o}m
solution [2], Kerr solution [3], and Kerr--Newman solution [4] are obtained.
These solutions describe static gravitational fields of an isolated body.
However, these solutions have a singularity at the origin which is believed
to be unavoidable. Due to the singularity the proper time at the origin
diverges and self force becomes infinite. So it seems inappropriate to
consider them as the gravitational fields of a point particle which can move
freely. To circumvent this problem there have been attempts to describe an
elementary particle as an extended object; the exterior Kerr--Newman metric
is matched to a nonsingular interior solution to the Einstein equations [5].
But such models are plagued by various unphysical features.
\par
For a system of many particles one cannot impose any symmetry conditions
because particles are forced to move under mutual interaction. Thus the
exact solution for many particle system is expected to be very complicated.
Nobody has succeeded in finding such exact solution for even the simplest
case of two particles. Instead one usually relies on approximation method by
expanding the metric tensor into a power series of a small parameter.
Einstein, Infeld, and Hoffmann [6] used $1/c$ as the small parameter and
obtained relativistic correction of order $1/{c^2}$ to the Newtonian theory.
Their method is known as slow approximation method since it is based on the
assumption that the time derivative of any field quantity is much smaller than
the spatial derivatives. But such an unsymmetrical treatment of time and space
coordinates goes against the spirit of the theory of relativity.
Havas and Goldberg [7] used the Newtonian gravitational constant $G$ as the
small parameter and obtained a Lorentz covariant approximation. This is known
as fast approximation method because it can be applied to the study of fast
moving bodies and also allows the investigation of gravitational radiation
damping. The method used in this work is analogous to this, but it is not
approximation.
\par
One more problem one has to take into account in the process of solving the
Einstein equations is the equations of motion. It is well known that the
equations of motion of a material body needs not be postulated; rather, they
follow from the field equations [8]. Thus, to solve the Einstein equations
one also has to solve the equations of motion. In the course of solving the
equations of motion, however, one expects that the proper time and the
Christoffel symbols become infinite at the location of the particle so that
the equations of motion contain infinite self--action terms and even may not
be consistent. When one solves the equations of motion in Newtonian theory,
one simply drops infinite self--action term by hand. But in general
relativity there is no way of dropping self--action terms consistently due
to the nonlinearity of the Christoffel symbols. Carmeli [9] explored the
possibility of obtaining the equations of motion which were free of infinite
self--action terms under certain assumptions. In this work we will show
that the self--action is zero if we choose a proper coordinate system.
\par
Another important feature of general relativity is arbitrariness in the
choice of coordinate system. In every text book in general relativity
arbitrariness in the choice of coordinate system is stressed from the
beginning. However, this arbitrariness raises a difficult problem of
distinguishing between real physical effects and coordinate effects.
Thus we are faced with an equivalence problem [10], namely the question
whether two given solutions to the Einstein equations describe the same
gravitational field. But the solution to this problem can given only locally,
and not globally. Due to this even if we have solved the Einstein equations,
we are still left with problem of fixing global topology. We will consider
this problem in conjunction with the Schwarzschild solution.
\par
In this work we consider the gravitational field of a moving point particle
as a first step toward the exact solution for many body problem in general
relativity. We avoid arbitrariness of coordinates by choosing a coordinate
system analogous to an inertial frame such that spacetime metric reduces
to Minkowski metric as the mass of the particle approaches zero.
Since the background spacetime is Minkowski space, we keep Lorentz covariance
at every stage of calculation. The idea is as follows: Suppose that we have an
exact solution to the Einstein equations. If we expand the metric tensor with
respect to the gravitational coupling constant, each term in the expansion is
a Lorentz covariant tensor.  Collecting the terms with the same Lorentz
tensors, we can express the metric tensor as a sum of Lorentz covariant terms
and each term in the sum as a product of a Lorentz invariant quantity and a
Lorentz covariant tensor. Since the numbers of Lorentz invariant quantities
and Lorentz covariant tensors are limited, the metric tensor can be expressed
in a simple form. Thus we make an ansatz for the metric tensor in a Lorentz
covariant form. Then substituting the form of the metric tensor into the
Einstein equations, we obtain a set of ordinary differential equations which
can be solved straightforward.
\par
This paper is organized as follows: In section 2 we solve the linearized
Einstein equations in a Lorentz covariant way in close analogy with
electromagnetism. We also investigate how the equations of motion follow
from the field equations by analyzing the linearized Einstein equations.
Section 3 describes the full nonlinear field equations. We make an ansatz for
the metric tensor from the knowledge of the linearized solution. Then
substituting the metric tensor into the Einstein equations which are originally
partial differential equations, we obtain a set of ordinary differential
equations. Solving these ordinary differential equations with proper Newtonian
limit, we obtain explicit form of the metric tensor for a moving point
particle. Using the exact solution, we examine the proper time at the
location of the particle and the self--action. We also consider an equivalence
problem for these solutions. Finally, in section 4 we discuss our results and
its implications. Notational conventions and some useful identities are listed
in appendix.
We use units in which $c=G=1$ and the flat spacetime metric $\ndd\mu\nu =
\text{diag}(-1,1,1,1)$.
\bigskip\flushpar
{\bf 2. Linearized equations}
\medskip\par
Before we consider the linearized Einstein equations, we first solve the
Maxwell's equations for a point charge in a Lorentz covariant way. Then we
analyze the linearized Einstein equations in close analogy with the Maxwell's
equations.
\bigskip\flushpar
{2.1 \it Electromagnetism}
\medskip\par
The Lagrangian density for the Maxwell fields in Gaussian unit can be written
as
$$ \Cal L=-\frac1{16\pi} F^{\mu\nu}F_{\mu\nu} \tag2.1 $$
and the field equations are
$$ \du 2 A_\mu - \dd\mu\du\nu A_\nu = -4\pi\,j_\mu. \tag2.2 $$
Since (2.2) is not directly invertible, we add a gauge fixing term
$$ \Cal L_{\text{GF}}=-\frac1{8\pi\xi}(\dr{\cdot}A)^2 \tag2.3 $$
to the Lagrangian density and obtain modified field equations
$$ \du 2 A_\mu - \bigl(1-\frac1\xi\bigr)\dd\mu\du\nu A_\nu = -4\pi\,j_\mu.
   \tag2.4 $$
Introducing Green's function which satisfies
$$ \du 2 G_{\mu\nu}(x) - \bigl(1-\frac1\xi\bigr)\dd\mu\du\lambda
   G_{\lambda\nu}(x) = -4\pi\,\ndd\mu\nu \delta(x), \tag2.5$$
we obtain the retarded Green's function
$$  G^{\text{ret}}_{\mu\nu}(x)=\theta(x^0)\,\bigl[(1+\xi)\ndd\mu\nu\,
    \delta(x^2) - 2(1-\xi)x_\mu x_\nu\,\delta'(x^2)\bigr]. \tag2.6 $$
We choose the gauge parameter $\xi=1$ so that the Green's function has the
simplest form
$$  G^{\text{ret}}_{\mu\nu}(x)=2\ndd\mu\nu\,\theta(x^0)\,
    \delta(x^2). \tag2.7 $$
\par
Now let a point charge $q$ move along the world line $\pos_0(t)$ with four
velocity $u^\mu(t)$. Then the current four vector is
$$ j^\mu=\frac q{\gamma(t)}\,u^\mu(t)\,\delta\bigl(\pos-\pos_0(t)\bigr).
   \tag2.8 $$
Using (2.7) and solving for $A_\mu$, we obtain the Li{\'e}nard--Wiechert
potential
$$ A_\mu=\frac{q\,u_\mu(t')}{\gamma(t')
   [1-\boldkey n(t'){\cdot}\boldkey v(t')]|\pos-\pos_0(t')|}, \tag 2.9 $$
where $t'$ is the retarded time given by
$$ t'=t-|\pos - \pos_0(t')| \tag2.10 $$
and  $\boldkey n$ is a unit vector
$$ \boldkey n(t') = \frac{\pos - \pos_0(t')}{|\pos - \pos_0(t')|}. \tag2.11 $$
Since the retarded time satisfies the light--cone constraint $t-t'=
|\pos-\pos_0(t')|$, we can define a null four vector $ R^\mu(t')$ with
components
$$ R^\mu(t')=(|\boldkey r-\boldkey r_0(t')|,\,\boldkey r-\boldkey r_0(t')).
   \tag2.12 $$
Taking scalar product of this null vector with the four velocity
$u^\mu(t')$ of the point charge, we obtain a Lorentz invariant quantity
$$ \ru=-\gamma(t')[1-\boldkey n(t'){\cdot}\boldkey v(t')]|\pos-\pos_0(t')|.
   \tag2.13 $$
Here the time variables in $R$ and $u$ on the left hand side are omitted. From
now on it should be understood that all the time variables omitted in $R^\mu$
and $u^\mu$ should be taken as the retarded time $t'$. Now $A_\mu$
can be expressed in a  simple Lorentz covariant form
$$ A_\mu = -\frac{q u_\mu}{\ru}. \tag2.14 $$
\par
Note that the Green's function $G^{\text{ret}}_{\mu\nu}(x)$ in (2.6) is
infinite for $\xi=\infty$ which correspond to the case before we add the gauge
fixing term to the Lagrangian density. Thus it may be possible that (2.14)
satisfies (2.4) with $\xi=1$ but not (2.2). But if we calculate the four
divergence of $A_\mu$ in (2.14), we find
$$ \dd\mu A^\mu=0 \tag2.15 $$
regardless of motion of the point charge. Hence the solution (2.14) also
satisfies the field equations (2.2). Therefore the equations of motion of the
point charge is independent of the field equations. We will see that this is
not the case with gravitational field.
\vfil\eject\flushpar
{2.2 \it Linearized gravity}
\smallskip\par
Now consider the field equations of the linearized gravity. The Lagrangian
density for the linearized gravity is given by
$$ \spreadlines{1\jot}
   \Cal L=\frac14(\du\lambda\huu\mu\nu)(\dd\lambda\hdd\mu\nu)
   -\frac14(\du\lambda h)(\dd\lambda h)
   -\frac12(\du\lambda \hdd\mu\lambda)(\dd\rho\huu\mu\rho)
   +\frac12(\du\lambda\hdd\mu\lambda)(\du\mu h), \tag2.16a $$
where
$$ h = \hdu\lambda\lambda. \tag2.16b $$
The field equations are
$$ \du 2\hdd\mu\nu - \dd\mu\du\lambda\hdd\lambda\nu
   - \dd\nu\du\lambda\hdd\mu\lambda
   + \ndd\mu\nu\du\lambda\du\rho\hdd\lambda\rho
   + \dd\mu\dd\nu h -\ndd\mu\nu\du 2h = -16\pi\,T_{\mu\nu}. \tag2.17 $$
Since these equations are not directly invertible, we add a gauge fixing term
$$ \Cal L_{\text{GF}}=\frac1{2\xi}\bigl(\dd\lambda\huu\mu\lambda
   - \frac12\du\mu h\bigr)\bigl(\du\rho\hdd\mu\rho-\frac12\dd\mu h\bigr)
   \tag2.18 $$
to the Lagrangian density and obtain modified field equations
$$ \align
   \du 2\hdd\mu\nu &-\bigl(1-\frac1\xi\bigr)(\dd\mu\du\lambda\hdd\lambda\nu
   + \dd\nu\du\lambda\hdd\mu\lambda)
   + \bigl(1-\frac1\xi\bigr)(\ndd\mu\nu\du\lambda\du\rho\hdd\lambda\rho
   + \dd\mu\dd\nu h)\\
   &-\bigl(1-\frac1{2\xi}\bigr)\ndd\mu\nu\du 2h =
   -16\pi\,T_{\mu\nu}. \tag2.19
   \endalign  $$
Introducing the Green's function which satisfies
$$ \aalign
   &\du 2G_{\mu\nu,\alpha\beta}(x)-\bigl(1-\frac1\xi\bigr)[\dd\mu\du\lambda
   G_{\lambda\nu,\alpha\beta}(x)+\dd\nu\du\lambda G_{\mu\lambda,
   \alpha\beta}(x)]\\
   &\quad+\bigl(1-\frac1\xi\bigr)[\ndd\mu\nu\du\lambda\du\rho
   G_{\lambda\rho,\alpha\beta}(x)+\dd\mu\dd\nu\eta^{\lambda\rho}
   G_{\lambda\rho,\alpha\beta}(x)]\\
   &\quad-\bigl(1-\frac1{2\xi}\bigr)\ndd\mu\nu
   \eta^{\lambda\rho}\du2G_{\lambda\rho,\alpha\beta}(x)
   = -\frac12(\ndd\mu\alpha\ndd\nu\beta+\ndd\mu\beta\ndd\nu\alpha)
   \delta(x), \tag2.20
   \endalign $$
we obtain the retarded Green's function
$$ \aalign
   G^{\text{ret}}_{\mu\nu,\alpha\beta}(x)&=\frac{\theta(x^0)}{4\pi}
   \bigl\{[\xi(\ndd\mu\alpha\ndd\nu\beta+\ndd\mu\beta\ndd\nu\alpha)-
   \ndd\mu\nu\ndd\alpha\beta]\,\delta(x^2)\\
   &\qquad - (1-\xi)(\ndd\mu\alpha x_\nu x_\beta +
   \ndd\mu\beta x_\nu x_\alpha + \ndd\nu\alpha x_\mu x_\beta +
   \ndd\nu\beta x_\mu x_\alpha)\,\delta'(x^2)\bigr\}. \tag2.21
   \endalign $$
We also choose the gauge parameter $\xi=1$ so that the retarded Green's
function takes the simplest form
$$ G^{\text{ret}}_{\mu\nu,\alpha\beta}(x)=\frac1{4\pi}
   (\ndd\mu\alpha\ndd\nu\beta+\ndd\mu\beta\ndd\nu\alpha-
   \ndd\mu\nu\ndd\alpha\beta)\,\theta(x^0)\,\delta(x^2). \tag2.22 $$
\par
Suppose that a point particle moves along the world line $\pos_0(t)$ with
four velocity $u^\mu(t)$. Then using the energy--momentum tensor
$$ T_{\mu\nu} = \frac m{\gamma(t)} u_\mu(t) u_\nu(t)\,
   \delta\bigl(\pos-\pos_0(t)\bigr) \tag2.23 $$
and solving for $\hdd\mu\nu$, we obtain the Li{\'e}nard--Wiechert potential
corresponding to spin 2 field
$$ \hdd\mu\nu=\frac{2m[\ndd\mu\nu + 2u_\mu(t') u_\nu(t')]}{\gamma(t')
   [1 - \boldkey n(t'){\cdot}\boldkey v(t')]|\pos - \pos_0(t')|}, \tag2.24 $$
or in terms of the Lorentz invariant quantity $\ru$ [7]
$$ \hdd\mu\nu=-\frac{2m(\ndd\mu\nu+2u_\mu u_\nu)}{\ru}. \tag2.25 $$
\par
Now note that the Green's function $G^{\text{ret}}_{\mu\nu,\alpha\beta}(x)$
in (2.21) is divergent for $\xi=\infty$. This means that the equations for the
Green's function are not invertible before we add the gauge fixing term to the
Lagrangian density. Thus (2.25), which is the solution to (2.19) with $\xi=1$,
may not satisfy (2.17), which are the equations we want to solve.
Substituting (2.25) into (2.17) to check this, we find
$$ \aalign
   &\du 2\hdd\mu\nu - \dd\mu\du\lambda\hdd\lambda\nu
   - \dd\nu\du\lambda\hdd\mu\lambda
   + \ndd\mu\nu\du\lambda\du\rho\hdd\lambda\rho
   + \dd\mu\dd\nu h -\ndd\mu\nu\du 2h\\
   &\qquad = 4m\biggl[\frac{\ra(1+\ra)}{\rup3}+\frac{\rw}{\rup2}\biggr]
   \ndd\mu\nu -\frac{4m(1+\ra)}{\rup3}(R_\mu a_\nu + R_\nu a_\mu)\\
   &\qquad\qquad - \frac{4m}{\rup2}(R_\mu w_\nu + R_\nu w_\mu) -
   \frac{4m}{\rup2}(u_\mu a_\nu + u_\nu a_\mu), \tag2.26
   \endalign $$
where $a^\mu$ and $w^\mu$ are the four acceleration and its derivative as
defined in appendix. Unless $a_\mu=0$, the solution (2.25) does not satisfy
the linearized field equations. For a single particle system one can always
choose a coordinate system such that $a_\mu=0$, which corresponds to the
equations of motion. For many particle system one can obtain the solution
for $\hdd\mu\nu$ by superposing (2.25), but one cannot choose a coordinate
system such that the accelerations of all particles are zero  simultaneously.
Thus to satisfy the linearized field equations (2.17), one has to add terms of
order $G^2$ and higher to $\hdd\mu\nu$ which, when substituted into (2.17),
will balance the terms on the right hand side of (2.26) and yield the
equations of motion. Hence the equations of motion in general relativity
follow from the field equations due to the kinematics of the Pauli--Fierz
equations.
\bigskip\flushpar
{\bf 3. Field equations}
\medskip\par
We now solve full nonlinear field equations for a moving point particle.
To avoid arbitrariness in coordinate system, we choose a coordinate system
analogous to an inertial frame such that spacetime metric reduces to
Minkowski metric as the mass of the source approaches zero. Since an isolated
object can not accelerate by itself, the four velocity of the particle must be
constant. Then the Lorentz covariant quantities we can use are $\ndd\mu\nu$,
$R_\mu$ and $u_\mu$, while the only Lorentz invariant quantity is $\ru$.
 From the solution of the linearized equation, we make an ansatz for the metric
tensor as
$$ \aalign
   \gdd\mu\nu&=e^A[\ndd\mu\nu+(1-e^B)\,u_\mu u_\nu], \tag3.1a\\
   \guu\mu\nu&=e^{-A}[\nuu\mu\nu+(1-e^{-B})u^\mu u^\nu],\tag3.1b\\
   \endalign $$
where $A$ and $B$ are functions of the Lorentz invariant quantity $\ru$ only.
The simple relation between $\gdd\mu\nu$ and $\guu\mu\nu$ follows from the
facts that $\ndd\mu\nu+u_\mu u_\nu$ is idempotent and that $\ndd\mu\nu +
u_\mu u_\nu$ and $u_\mu u_\nu$ are orthogonal. We also note that the metric
(3.1) is similar to Kerr--Schild metric which allows one to use Lorentz
invariance in a general relativistic context [11]. After straightforward
calculation we obtain the Christoffel symbol
$$ \aalign
   \Gamma^\lambda_{\mu\nu}&=\frac{A'}2(\delta^\lambda_\nu u_\mu+
   \delta^\lambda_\mu u_\nu-\ndd\mu\nu u^\lambda)
   -\bigl\{B'+\frac12[A'-(A'+B')e^B]\bigr\}u^\lambda u_\mu u_\nu\\
   &\qquad+\frac{A'}{2\ru}(\delta^\lambda_\nu R_\mu+\delta^\lambda_\mu
   R_\nu-\ndd\mu\nu R^\lambda)-\frac{B'}{2\ru}u^\lambda(R_\mu u_\nu+
   R_\nu u_\mu)\\
   &\qquad-\frac{[A'-(A'+B')e^B]}{2\ru}R^\lambda u_\mu u_\nu, \tag3.2
   \endalign $$
and the Ricci tensor
$$ \aalign
   R_{\mu\nu}&=-\frac12\biggl(A''+\frac4{\ru}A'+\frac1{\ru}B'+{A'}^2+
   \frac12A'B'\biggr)\ndd\mu\nu\\
   &\quad-\frac1{\rup2}\biggl(A''-\frac1{\ru}A'+\frac12B''-\frac1{2\ru}B'-
   \frac12{A'}^2+\frac14{B'}^2\biggr)R_\mu R_\nu\\
   &\quad-\frac1{\ru}\biggl(A''-\frac1{\ru}A'+\frac12B''-\frac1{2\ru}B'-
   \frac12{A'}^2+\frac14{B'}^2\biggr)(R_\mu u_\nu+u_\mu R_\nu)\\
   &\qquad+\biggl\{-\frac32A''-\frac12B''+\frac12(A''+B'')e^B-\frac1{\ru}[A'-
   (A'+B')e^B]\\
   &\hphantom{\qquad+\biggl\{}\quad-\frac14A'B'-\frac14{B'}^2+\frac14(2A'+B')
   (A'+B')e^B\biggr\}u_\mu u_\nu. \tag3.3
   \endalign $$
We now solve the Einstein equations for both uncharged and charged cases.
\bigskip\flushpar
{3.1 \it Uncharged case}
\medskip\par
The energy--momentum tensor of a point particle is
$$ T^{\mu\nu} = \frac m{\sqrt{-g}}\frac{d\tau}{dt} u^\mu u^\nu
   \delta\bigl(\pos - \pos_0(t)\bigr). \tag3.4 $$
So except at the location of the particle the Einstein equations become
$ R_{\mu\nu} = 0$. Thus
$$ \aalign
   &A''+\frac4{\ru}A'+\frac1{\ru}B'+{A'}^2+\frac12A'B'=0, \tag3.5a\\
   &A''-\frac1{\ru}A'+\frac12B''-\frac1{2\ru}B'-\frac12{A'}^2+\frac14{B'}^2
   =0, \tag3.5b\\
   -&\frac32A''-\frac12B''+\frac12(A''+B'')e^B-\frac1{\ru}[A'-(A'+B')e^B]\\
   &\hphantom{\frac32A''}-\frac14A'B'-\frac14{B'}^2+\frac14(2A'+B')(A'+B')e^B
   =0. \tag3.5c
   \endalign $$
Combining these equations, we have
$$ A''+\frac2{\ru}A'+\frac14{A'}^2=0, \tag3.6a $$
$$ \biggl(\frac32A+B\biggr)''+\frac2{\ru}\biggl(\frac32A+B\biggr)'
   + \frac12\biggl(\frac32A'+B'\biggr)^2=0. \tag3.6b $$
The solutions to these equations with proper Newtonian limit are
$$ A=4\ln\left(1-\frac m{2\ru}\right), \tag3.7a $$
$$ \frac32A+B=2\ln\left(1+\frac m{2\ru}\right). \tag3.7b $$
Therefore
$$ \gdd\mu\nu =\biggl(1-\frac m{2\ru}\biggr)^4(\ndd\mu\nu + u_\mu u_\nu) -
   \left(\frac{1+\dfrac m{\mathstrut 2\ru}}{1-\dfrac{\mathstrut m}
   {2\ru}}\right)^2 u_\mu u_\nu.\tag3.8 $$
\par
The proper time used in definition of the four velocity of the source is
special relativistic one. Since we have solved the full nonlinear equations,
we now consider general relativistic one. Since the proper time interval at
the location of the particle is
$$ d\tau = \frac1\gamma[-\gdd\mu\nu u^\mu u^\nu]^{1/2}_{\vphantom{q}_{R=0}}\,
   dt = \frac1\gamma\, dt , \tag3.9 $$
it is the same as the special relativistic one. This could have been noticed
earlier if we had taken a close look at the Schwarzschild solution in
isotropic coordinates, which is the static limit of the solution we obtained
with the particle at the origin.
$$ ds^2 = -\left(\frac{1-\dfrac m{\mathstrut 2r}}{1+\dfrac{\mathstrut m}{2r}}
   \right)^2 dt^2 + \biggl(1+\dfrac m{2r}\biggr)^4 (dx^2 + dy^2 + dz^2).
   \tag3.10 $$
In this coordinate system the proper time interval at $r=0$ is the same as the
coordinate time interval.
\par
We can transform this expression to usual form
$$ ds^2 = -\biggl(1-\frac{2m}{\rb}\biggr)dt^2 + \frac{d\rb^2}
   {1-\dfrac{\mathstrut 2m}{\rb}} + \rb^2(d\theta^2 + \sin^2\theta\,d\phi^2)
   \tag3.11 $$
by introducing a new radius variable $\rb$
$$ \rb=\biggl(1+\frac m{2r}\biggr)^2 r, \tag3.12a $$
or
$$ r=\frac12[\rb-m \pm ({\rb}^2-2m \rb)^{1/2}]. \tag3.12b $$
Since the nonlinear transformation (3.12) is not one--to--one,
the region between $\rb=2m$ and $\rb=\infty$ is mapped twice, while the region
between $\rb=0$ and \hbox{$\rb=2m$} is not mapped by any value of $r$. So the
nonlinear transformation (3.12a) induces a nontrivial topology in
Schwarzschild coordinates. To visualize the global topology of spacetime in
Schwarzschild coordinates, let us consider embedding of a spacelike
hypersurface of constant time $t=0$, with one degree of rotational freedom
suppressed ($\theta=\pi/2$). In Schwarzschild coordinates this hypersurface
corresponds to two distinct asymptotically flat hypersurfaces which are
connected by the Einstein--Rosen bridge [12] at $\rb=2m$. Now let a test
particle move from $r=\infty$ to $r=0$ in isotropic coordinate and
consider the trajectory of its image in Schwarzschild coordinates.
As it moves from $r=\infty$ to $r=m/2$, its image moves from
$\rb=\infty$ to $\rb=2m$ in one of the two hypersurfaces. As it crosses the
circle $r=m/2$ and moves to $r=0$, its image switches the other hypersurface
by crossing the Einstein--Rosen bridge at $\rb=2m$ and goes to $\rb=\infty$.
Thus the location of the particle $r=0$ is pushed to $\rb=\infty$.
This is the reason why one couldn't find nonstatic
generalization of the Schwarzschild solution in Schwarzschild coordinates.
\par
To check the consistency of the solution (3.8), we consider the gravitational
self force and find
$$ F_{\text{grav}}^\lambda =-m[\Gamma^\lambda_{\mu\nu}]_{\vphantom{q}_{R=0}}
   u^\mu u^\nu =0 \tag3.13 $$
This is consistent with constancy of the four velocity.  This can be inferred
from the fact that the location of the particle $r=0$ corresponds to
$\rb=\infty$ where the metric tensor reduces to that of flat spacetime.
\bigskip\flushpar
{3.2 \it Charged case}
\medskip\par
 From (2.14), we make an ansatz for the four vector potential of a point
charge as
$$ A_\mu = C u_\mu, \tag3.14 $$
where $C$ is a function of Lorentz invariant quantity $\ru$ only. Then the
field strength tensor is
$$ F_{\mu\nu}=\frac{C'}{\ru}(R_\mu u_\nu - R_\nu u_\mu) \tag3.15 $$
and the Maxwell's equations $ F^{\mu\nu}{}_{|\nu}=0 $ become
$$ [\rup2 e^{-B/2}C']'=0. \tag3.16 $$
Since $C=-q/{\ru}$ in special relativistic limit , we have
$$ C'=\frac{q\,e^{B/2}}{\rup2}. \tag3.17 $$
Substituting (3,17) into (3.15) and calculating the energy--momentum tensor
of electromagnetic field, we have
$$ \aalign
   T_{\mu\nu} &=\frac{q^2 e^{-A}}{8\pi\rup4}\biggl[\ndd\mu\nu -
   \frac2{\rup2}R_\mu R_\nu - \frac2{\ru}(R_\mu u_\nu + R_\nu u_\mu)\\
   &\hphantom{=\frac{q^2 e^{-A}}{8\pi\rup4}\biggl[\ndd\mu\nu } +
   \frac12(e^B-1)u_\mu u_\nu\biggr]. \tag3.18
   \endalign $$
Since $T_{\mu\nu}$ is traceless, the Einstein equations can be written as
$R_{\mu\nu}=8\pi T_{\mu\nu}$. Thus
$$ \aalign
   &A''+\frac4{\ru}A'+\frac1{\ru}B'+{A'}^2+\frac12A'B'=-\frac{2q^2 e^{-A}}
   {\rup4}, \tag3.19a\\
   &A''-\frac1{\ru}A'+\frac12B''-\frac1{2\ru}B'-\frac12{A'}^2+\frac14{B'}^2
   = \frac{2 q^2 e^{-A}}{\rup4}, \tag3.19b\\
   -&\frac32A''-\frac12B''+\frac12(A''+B'')e^B-\frac1{\ru}[A'-(A'+B')e^B]
   -\frac14A'B'\\
   &\hphantom{\frac32A''}-\frac14{B'}^2+\frac14(2A'+B')(A'+B')e^B=
   \frac{q^2 e^{-A}(e^B-1)}{\rup4}. \tag3.19c
   \endalign $$
Combining these equations, we obtain
$$ A''+\frac2{\ru}A'+\frac14{A'}^2=-\frac{q^2 e^{-A}}{\rup4}, \tag3.20a $$
$$ \biggl(\frac32A+B\biggr)''+\frac2{\ru}\biggl(\frac32A+B\biggr)'
   +\frac12\biggl(\frac32A'+B'\biggr)^2=\frac{3q^2 e^{-A}}{2\rup4}.\tag3.20b $$
The solutions to these equations which reduce to (3.7) as $q\rightarrow 0$ are
$$ A=2\ln\biggl[1-\frac m{\ru} + \frac{m^2-q^2}{4\rup2}\biggr],
   \tag3.21a $$
$$ \frac32A+B=2\ln\biggl[1-\dfrac{m^2 - q^2}{4\rup2}\biggr] -
   \ln\biggl[1-\dfrac m{\ru} + \dfrac{m^2-q^2}{4\rup2}\biggr].
   \tag3.21b $$
Therefore
$$ \gdd\mu\nu =\biggl[1-\frac m{\ru}+\frac{m^2-q^2}{4\rup2}\biggr]^2
   (\ndd\mu\nu + u_\mu u_\nu) - \frac{\biggl[1-\dfrac{m^2 - q^2}{4\rup2}
   \biggr]^2}{\biggl[1-\dfrac m{\ru} + \dfrac{m^2-q^2}{4\rup2}\biggr]^2}\,
   u_\mu u_\nu.\tag3.22 $$
Substituting (3.21) into (3.17) and integrating, we obtain
$$ C= -\frac{\dfrac q{\ru}}{1 - \dfrac m{\ru} + \dfrac{m^2-q^2}{4\rup2}}.
   \tag3.23 $$
Hence
$$ A_\mu= -\frac{\dfrac {q\,u_\mu}{\ru}}{1 - \dfrac m{\ru} +
   \dfrac{m^2-q^2}{4\rup2}}. \tag3.24 $$
\par
As for uncharged case the gravitational self force is zero. Considering
the electromagnetic self force, we also find
$$ F_{\text{em}}^\lambda = q[F^\lambda{}_\sigma u^\sigma]_{\vphantom{q}_{R=0}}
   =0. \tag 3.25 $$
The proper time interval at the location of the particle is the same as the
special relativistic one except for the extreme cases $q=\pm m$. In the cases
$q=\pm m$ it is interesting to note that the proper time interval at the
location of the particle is zero even if mass of the particle is not zero.
In this case one can choose the special relativistic proper time as the Affine
parameter for the trajectory of the particle and the equations of motion are
still consistent. In the case of many point sources with $q_i=\pm m_i$
the electrostatic repulsions exactly balance the gravitational attractions.
Majumdar [13] and Papapetrou [14] discovered a static solution to the
Einstein--Maxwell equations corresponding to this case. In the limit of a
single point source their solution reduces to the static limit of (3.22) with
$q=\pm m$.
\bigskip\flushpar
{\bf 4. Discussion}
\medskip\par
We have developed a Lorentz covariant method of solving the Einstein equations
and have obtained the exact solution to the Einstein equations with a moving
point particle in a Lorentz covariant form for both uncharged and charged
cases. Analyzing the linearized Einstein equations, we also have found that the
equations of motion follow from the field equations due to the structure of
the linearized Einstein equations. It is shown that the general relativistic
proper time interval at the location of the particle is the same as the special
relativistic one and the gravitational and electromagnetic self forces are
zero. Hence one doesn't have to worry about how to drop self--action term by
hand, if one has chosen a proper coordinate system.
This is the power of the exact solution which cannot be obtained by any
kind of approximation method.
\par
In the course of examining the solution we encountered an equivalence problem.
As we have seen, nonlinear coordinate transformation may
induce distortion of topology. Since the Einstein equations determine only
the local geometry of spacetime and not its topology , one has to
choose a coordinate system with proper topology which meets a certain
criterion. The criterion we used is possibility of consistent description of
the source as a point particle. Therefore the most natural coordinate system
for a spherically symmetric point source is the
isotropic coordinates rather than the Schwarzschild coordinates.
\par
One may extend the method used in this work to obtain exact solution to
the Einstein equations with a moving spinning point particle. This will be
presented in a separate publication.
\vfil\eject
{\bf Acknowledgement}
\medskip\par
I would like to thank Professor Robert Finkelstein for reading the manuscript
and comments and Professors Young Jik Ahn and Taejin Lee for helpful
discussions.
\bigskip\flushpar
{\bf Appendix. Conventions and useful identities}
\medskip\par
The special relativistic four velocity, four acceleration, and its derivative
of a point particle moving along the world line $x^\mu(t)$ are defined by
$$ \align
   u^\mu(t)&=\gamma(t)\drt{x^\mu}t, \tag A.1a\\
   a^\mu(t)&=\gamma(t)\drt{u^\mu}t, \tag A.1b\\
   w^\mu(t)&=\gamma(t)\drt{a^\mu}t, \tag A.1c
   \endalign $$
where
$$ \gamma =\frac1{\sqrt{1-v^2}}. \tag A.1d $$
They satisfy
$$ \aalign
   &\dd\mu R_\nu = \ndd\mu\nu - \frac{R_\mu u_\nu}{\ru}, \tag A.2a\\
   &\dd\mu u_\nu = \frac{R_\mu a_\nu}{\ru}, \tag A.2b\\
   &\dd\mu a_\nu = \frac{R_\mu w_\nu}{\ru}, \tag A.2c
   \endalign $$
where all the time variables omitted in $R$, $u$, $a$ and $w$ are the retarded
time $t'$.
\par
Let the metric tensor be of the form
$$ \gdd\mu\nu=\ndd\mu\nu + \hdd\mu\nu, \tag A.3 $$
then the determinant of the metric tensor and the contravariant metric tensor
can be written in Lorentz covariant forms as
$$ \aalign
   -g&=1+\tr h+\frac12\bigl[(\tr h)^2-\tr h^2\bigr]
   +\frac16\bigl[(\tr h)^3 -3\,\tr h\,\tr h^2+2\,\tr h^3\bigr]\\
   &\qquad +\frac1{24}\bigl[(\tr h)^4-6(\tr h)^2\,\tr h^2 +
   8\,\tr h\,\tr h^3 + 3(\tr h^2)^2-6\,\tr h^4\bigr], \tag A.4
   \endalign $$
\vfil\eject
$$ \aalign
   g^{\mu\nu}&=-\frac1g\biggl\{\bigl[1+\tr h+\frac12(\tr h)^2-\frac12\tr h^2
   +\frac16(\tr h)^3-\frac12\tr h\,\tr h^2+\frac13\tr h^3\bigr]\,
   \nuu\mu\nu\\
   &\hphantom{-\frac1g\biggl\{}\qquad-\bigl[1+\tr h+\frac12(\tr h)^2 -
   \frac12\tr h^2\bigr]\,\huu\mu\nu + (1+\tr h)\,\huu\mu\lambda\,
   \hdu\lambda\nu\\
   &\hphantom{-\frac1g\biggl\{}\qquad-\huu\mu\lambda\,\hdd\lambda\sigma\,
   \huu\sigma\nu\biggr\}, \tag A.5
   \endalign $$
where
$$ \tr h= \hdu\mu\mu, \tag A.6a $$
$$ \tr h^2= \hdu\mu\nu\hdu\nu\mu, \tag A.6b $$
$$ \tr h^3= \hdu\mu\nu\hdu\nu\lambda\hdu\lambda\mu, \tag A.6c $$
$$ \tr h^4= \hdu\mu\nu\hdu\nu\lambda\hdu\lambda\sigma\hdu\sigma\mu.\tag A.6d $$
Here we used the flat spacetime metric when raising or lowering indices in $h$.
\bigskip\flushpar
{\bf References}
\medskip\flushpar
\item{[1]} Schwarzschild K 1916 {\it Sitzber\. Deut\. Akad\. Wiss\. Berlin
Kl\. Math\. Phys\. Tech\.} 189
\item{[2]} Reissner H 1916 {\it Ann\. Physik} {\bf 50} 106 \newline
Nordstr{\"o}m G 1918 {\it Proc\. Kon\. Ned\. Akad\. Wet\.} {\bf 20} 1238
\item{[3]} Kerr R P 1963 {\it Phys\. Rev\. Lett\.} {\bf 11} 237
\item{[4]} Newman E T, Couch E, Chinnapared K, Exton A, Prakash A and
Torrence R 1965 {\it J\. Math\. Phys\.} {\bf 6} 918
\item{[5]} L{\'o}pez C A 1984 {\it Phys\. Rev\.} {\bf D30} 313 \newline
Mc Manus D 1991 {\it Class\. Quant\. Grav\.} {\bf 8} 863
\item{[6]} Einstein A, Infeld L and Hoffmann B 1938 {\it Ann\. Math\.} {\bf 39}
65
\item{[7]} Havas P 1957 {\it Phys\. Rev\.} {\bf 108} 1351 \newline
Havas P and Goldberg J N 1962 {\it Phys\. Rev\.} {\bf 128} 398
\item{[8]} Einstein A and Grommer J 1927 {\it Sitzber\. Deut\. Akad\. Wiss\.
Berlin Kl\. Math\. Phys\. Tech\.} 2; 235
\item{[9]} Carmeli M 1965 {\it Phys\. Rev\.} {\bf 140} B1441
\item{[10]} Karlhelde A 1980 {\it Gen\. Rel\. Grav\.} {\bf 12} 693
\item{[11]} G{\"u}rses M and G{\"u}rsey F 1975 {\it J\. Math\. Phys\.}
{\bf 16} 2385 \newline
Einstein S and Finkelstein R 1977 {\it Phys\. Rev\.} {\bf D15} 2721
\item{[12]} Einstein A and Rosen N 1935 {\it Phys\. Rev\.} {\bf 48} 73
\item{[13]} Majumdar S D 1947 {\it Phys\. Rev\.} {\bf 72} 930
\item{[14]} Papapetrou A 1947 {\it Proc\. Roy\. Irish Acad\.} {\bf A51} 191
\enddocument